# Anisotropic properties of the layered superconductor $Cu_{0.07}TiSe_2$

E. Morosan[1], Lu Li[2], N. P. Ong[2], and R. J. Cava[1]

[1]Department of Chemistry, Princeton University, Princeton, NJ 08540, USA

[2]Department of Physics, Princeton University, Princeton, NJ 08540, USA

The anisotropic superconducting properties of single crystals of $Cu_{0.07}TiSe_2$ were studied by measurements of magnetization and electrical resistivity. $T_C$ is around 3.9 K, and the measured upper critical field ($H_{c2}$) values are ~1.25 T and 0.8 T, for applied field parallel and perpendicular to the $TiSe_2$ planes, respectively. The anisotropy ratio $\gamma_{anis} = H^{ab}_{c2} / H^{c}_{c2}$ is close to 1.6 and nearly temperature independent. The lower critical field ($H_{c1}$) values are much smaller (~ 32 Oe for H||ab and 17 Oe for H//c); demagnetizing corrections for field perpendicular to the thin plate crystals are required for the determination of $H^{c}_{c1}$. The anisotropy of the critical fields is described well by the anisotropic Ginzburg-Landau (GL) theory, and the characteristic GL parameters are determined and discussed.

**Introduction**

TiSe$_2$ is one of the first known charge density wave (CDW) compounds [1-5], with the CDW transition having been studied extensively due to its controversial nature. The crystal structure is that of the classical layered dichalcogenide type with hexagonal TiSe$_2$ layers stacked perpendicular to the crystallographic *c* axis. Cu intercalation between the TiSe$_2$ layers [6] has been recently shown to drive the CDW transition down in temperature and, at intermediate compositions (x ≥ 0.04 in Cu$_x$TiSe$_2$), to give rise to a new superconducting (SC) state. After the CDW transition is fully suppressed for x ≥ 0.06, the superconducting transition reaches a maximum value around x = 0.08. This implies a competition between the two collective electron states (CDW and SC) that remains to be further elucidated.

Before the connection between the CDW and the SC states in Cu$_x$TiSe$_2$ can be understood, characterization of the SC state is of interest. Here we report studies of plate-like single crystals of Cu$_{0.07}$TiSe$_2$, allowing us to investigate the anisotropic physical properties of this system via magnetization and transport measurements. For H||*ab*, the temperature dependence of the critical fields H$_{c1}$(T) and H$_{c2}$(T) are determined in a straightforward way. When field is applied perpendicular to the thin crystal plates (along the *c* axis), demagnetizing effects are considerable and must be taken into account. The upper critical field H$^c_{c2}$ is almost unaffected by demagnetization, but the much smaller measured value of the lower critical field H$^c_{c1}$ is significantly reduced from the effective field H$^{c*}_{c1}$ in this orientation. After demagnetizing field corrections are applied, the anisotropic parameters characteristic of the superconducting state are determined and analyzed within the context of the anisotropic Ginzburg-Landau theory.

**Experiment**

Single crystals of Cu$_{0.07}$TiSe$_2$ were grown via chlorine vapor transport. Polycrystalline TiSe$_2$ powders were first synthesized by mixing stoichiometric amounts of Ti and Se powders and then heating them up to 650°C at a rate of ~50°/hr in an evacuated silica tube. Next, CuCl$_2$ and TiSe$_2$ powders in a 0.10:1 ratio were sealed in 150 mm long evacuated silica tubes with a 12 mm diameter. The tubes were placed in a gradient furnace with the hot temperature set at 650°C and the cold end temperature kept at 550°C. After fourteen days, the furnace was cooled to room temperature over a few hours. Large hexagonal plates were formed towards the hot end.

X-ray diffraction measurements were employed to characterize the samples. Room temperature data were recorded on a Bruker D8 diffractometer using Cu Kα radiation and a graphite diffracted beam monochromator. Magnetization measurements as a function of temperature and applied field M(H,T) were performed in a Quantum Design MPMS SQUID magnetometer (T = 1.8 - 350 K, $H_{max}$ = 5.5 T). Anisotropic resistivity ρ(T,H) measurements with current parallel to the ab-plane were taken in a $^3$He refrigerator inserted in a 7 T superconducting magnet using a standard four probe technique.

**Results**

X-ray diffraction measurements on single crystals of $Cu_{0.07}TiSe_2$ showed them to have the structure previously reported on the polycrystalline materials. The copper content of 0.07 ± 0.01 per formula unit was determined by the *c* lattice parameter calibration [6] and was consistent with the observed $T_c$ of 3.9 K.

Fig. 1 shows the H = 0 resistivity data for $Cu_{0.07}TiSe_2$ as a function of temperature for current flowing in the *ab*-plane. At high temperatures, the resistivity is metallic in nature with ρ(T) increasing almost linearly with T. Upon lowering the temperature, a superconducting transition occurs at $T_c$ = 3.9 K, as determined from the maximum slope of the resistivity around the transition (Fig. 1, upper inset). The 10-90% width of the resistive transition in zero field is 0.1 K. Just above the transition, the resistivity shows the quadratic behavior ρ(T) = $ρ_0$ + A $T^2$ (Fig. 1, lower inset) expected for a Fermi liquid, with a residual resistivity $ρ_0$ = 80 μΩ cm and a coefficient A $\approx$ 1.1*$10^{-2}$ μΩ cm $K^{-2}$. $ρ_0$ and ρ(300K) are relatively large, about 80 μΩ cm and 520 μΩ cm respectively, and the residual resistivity ratio RRR = ρ(300K)/ρ(4.5K) is approximately 6.5. By comparison, superconducting layered $NbSe_2$ [7-8] shows smaller ρ(i//*ab*) values (between 2 and 100 μΩ cm for 8 K ≤ T ≤ 300 K), resulting in a RRR of approximately 40. We speculate that the intercalation of Cu atoms between the planes enhances the i∥*ab* scattering processes compared to those observed in undoped layered structures (e.g., $NbSe_2$).

Anisotropic measurements of the field dependent resistivity around the superconducting transition were performed down to $^3$He temperatures. These results are shown in Fig. 2, for field applied parallel or perpendicular to the hexagonal plates. (The $TiSe_2$ planes are in the planes of the crystal plates.) In both cases, the current was parallel to the *ab*-plane. For H||*ab* (Fig. 2a) the T = 0.35 K ρ(H) curve yields an upper critical field value $H^{ab}_{c2}$ = 1.22 T, as determined from the transition onset. An inflexion in the ρ(H) data is observed around H = 1.3 T, which moves lower in field as temperature increases. The origin of this feature is not known, but it may be associated with small inhomogeneities in the Cu doping concentration. The temperature dependence of $H^{ab}_{c2}$ is characteristic of a type-II superconductor, as the transition moves to lower and lower field values upon increasing the temperature. Very similar behavior is observed for the other field orientation (H//*c*, Fig. 2b) with smaller $H^{c}_{c2}$ values. At T = 0.35 K, $H^{c}_{c2}$ = 0.71 T, thus yielding an anisotropy ratio $\gamma_{anis} = H^{ab}_{c2} / H^{c}_{c2}$ in $Cu_{0.07}TiSe_2$ of about 1.7.

Fig. 3 shows the zero-field cooled M(H) isotherms for field H||*ab* (Fig. 3a) and H//*c* (Fig. 3b). These curves confirm the anisotropic suppression of the superconducting state observed in the magnetoresistance data. The minimum temperature available for these measurements is 1.8 K, where the H||*ab* magnetization becomes zero around $H^{ab}_{c2}(1.8K) \approx 0.96$ T. At the same temperature, the upper critical field $H^{c}_{c2}$ in the *c* direction is determined to be 0.54 T. The insets in Fig. 3 show the magnetization in low magnetic fields from which the $H_{c1}$ values can be estimated. For H||*ab* (inset, Fig. 3a) the M(H) data is linear for very low fields (dotted line). $H^{ab}_{c1}$ is determined as the point of departure from linearity, and these values are marked by vertical arrows for various temperatures. The analogous values are even smaller for H//*c* (inset, Fig. 3b) resulting in less accurate measurements of the magnetization values at such low fields. This allows us only to estimate the $H^{c}_{c1}$ value at T = 1.8 K (vertical arrow) as ~ 13 Oe.

Based on our resistivity and magnetization measurements, the anisotropic properties of the superconducting state in $Cu_{0.07}TiSe_2$ can be summarized in a $H_{c2}$ – T phase diagram (Fig. 4a): the full symbols have been determined either from the ρ(H) data (triangles) or M(H) curves (circles) for H||*ab*, and the open symbols correspond to the H//*c* direction. The inset shows the lower critical field values $H_{c1}$ as determined from the M(H) measurements. The dotted and dashed lines represent fits in the T → 0 and T → $T_c$ regions, respectively. An almost temperature-independent anisotropy ratio $\gamma_{anis} = H^{ab}_{c2} / H^{c}_{c2}$ is observed (Fig. 4b).

**Discussion**

As seen in Fig. 1, the H = 0 resistivity data of $Cu_{0.07}TiSe_2$ shows a superconducting transition around $T_c$ = 3.9 K. Above the transition, Fermi liquid behavior is observed, characterized by quadratic dependence of the resistivity on temperature: $\rho(T) = \rho_0 + A T^2$. The resistivity coefficient A is determined to be $\sim 1.1*10^{-2}$ $\mu\Omega$ cm/K$^2$. Using the H = 0 electronic specific heat coefficient $\gamma$ = 4.3 mJ/mol K$^2$ [6], the Kadowaki-Woods (KW) ratio A/$\gamma^2$ is estimated to be $\sim 60*10^{-5}$ $\mu\Omega$ cm/(mJ/mol K)$^2$. For many correlated electron systems, the Kadowaki-Woods ratio has been shown empirically to have a nearly universal value $a_0 = 10^{-5}$ $\mu\Omega$ cm/(mJ/mol K)$^2$ (solid line, Fig. 6) [9]. A number of compounds have since been shown to follow on almost universal curves with either reduced (dotted line, Fig. 6) or enhanced (dashed lines, Fig. 6) KW ratios [10-11]. It can be seen that $Cu_{0.07}TiSe_2$ falls close to the 60$a_0$ line, with many intermetallic compounds falling in-between the $a_0$ and the 60$a_0$ lines. Various mechanisms have been proposed to explain the former universality class, such as intersite magnetic correlations or ground state degeneracy. In $Na_xCoO_2$ [11] the KW ratio was found to be almost 50$a_0$, which has been attributed to magnetic frustration, proximity to a magnetic quantum critical point or a Mott transition. In the case of $Sr_2RuO_4$ [12], the KW ratio was found to be highly anisotropic, reaching values around 300$a_0$ for current normal to the $RuO_2$ planes; this was attributed to the two-dimensional character of the Fermi liquid in this compound. Since the thin-plate geometry of the $Cu_{0.07}TiSe_2$ precluded us from performing transport measurements with current perpendicular to the $TiSe_2$ planes, we can only speculate that the Fermi liquid might have a two-dimensional character, possibly inducing the high KW ratio. Another possibility is the proximity of this compound to the transition from a charge density (CDW) to a superconducting (SC) state, which may have significant influence on the scattering mechanism and, consequently, the KW ratio.

Next, we analyze the superconducting state in $Cu_{0.07}TiSe_2$. The BCS theory of superconductivity [13] predicts that, for T → $T_c$, the critical field behavior is almost linear in temperature. This is found for both field orientations for $H_{c2}$ (Fig. 4a) in $Cu_{0.07}TiSe_2$ as well as for the H∥ab $H_{c1}$ values (inset, Fig. 4a) and emphasized by the dotted lines. As described previously, the small $H^c_{c1}$ values correspond to very small measured magnetization values, making the accurate determination of the lower critical field difficult for this field orientation. In consequence, we are using the $H^c_{c1}$, determined at T = 1.8 K, and $T_{c,H=0}$ = 3.9 K, as determined from the in-plane data, to construct a dotted line analogous to that for H∥ab (inset, Fig. 4a). From the linear fits in the H = 0 limit, the critical temperature is determined to be $T_c$ = 3.9 K; towards

T = 0, the extrapolations of these linear fits can be used to estimate the $H_{c2}(0)$ critical field values by using the Werthamer-Helfand-Hohenberg (WWH) equation [14]:

$$H_{c2}(0) = 0.693 \cdot [-(dH_{c2}/dT)]_{T_c} \cdot T_c.$$

Since the lower critical field values are only determined from magnetization measurements limited to temperatures above 1.8 K, we also use the WWH formula to estimate $H_{c1}$ closer to T = 0. The $H_{c1}$ and $H_{c2}$ values thus determined are listed in Table I: $H^{ab}_{c1}(0) \approx 32$ Oe, $H^{c}_{c1}(0) \approx 17$ Oe, $H^{ab}_{c2}(0) \approx 1.23$ T and $H^{c}_{c2}(0) \approx 0.73$ T. The anisotropy ratio of $H_{c2}$, calculated as $\gamma_{anis}(0) = H^{ab}_{c2}(0) / H^{c}_{c2}(0)$ [15] is also listed in Table I. In addition, the temperature dependence of $\gamma_{anis}$ is shown in Fig. 4b. For most of the superconducting state, $\gamma_{anis}$ is almost constant and close to 1.63, except very close to $T_c$ where this value appears to be slightly higher; in this temperature range $H_{c1}$ and $H_{c2}$ for both field orientations approach zero, and thus $\gamma_{anis}$ is calculated as the ratio of two small quantities, and its determination is less certain. It can be concluded that the anisotropy $\gamma_{anis}$ is intrinsically temperature independent.

Close to T = 0, the upper critical field decreases with temperature as

$$H_{c2}(T) \approx H_{c2}(0) [1 - 1.07 (T/T_c)^2]. \; [13]$$

Fits to this expression are shown in Fig. 4a as dashed lines for both field orientations; for H||*ab*, the resulting critical parameters are $H^{ab}_{c2} = 1.24$ T and $T_c = 3.6$ K, while for H//*c*, $H^{c}_{c2} = 0.8$ T and $T_c = 3.4$ K. The estimated upper critical field values are close to those obtained before using the WWH formula, with the critical temperatures slightly smaller than the $T_c = 3.9$ K determined experimentally and from the linear fits.

The Ginzburg-Landau (GL) coherence length along the *i* direction $\xi_i$ is estimated from the anisotropic Ginzburg-Landau formulas [16] for $H_{c2}$: $H^{ab}_{c2} = \Phi_0/(2\pi \xi_{ab} \xi_c)$ and $H^{c}_{c2} = \Phi_0/(2\pi \xi^2_{ab})$, where $\Phi_0$ is the flux quantum $\Phi_0 = 2.07 \; 10^{-7}$ G cm$^{-2}$. The anisotropic values of the coherence length $\xi_{ab}(0)$ and $\xi_c(0)$ are listed in Table I. The T = 0 critical field values can also be used to determine the GL parameter $\kappa_i(0)$ along the i direction, using the equation $H^{i}_{c2}(0)/H^{i}_{c1}(0) = 2 \kappa^2_i(0) / \ln \kappa_i(0)$. In turn, the GL parameter $\kappa_i(0)$ is related to the coherence length $\xi_i(0)$ and the GL penetration depth $\lambda_i(0)$ as $\kappa_c(0) = \lambda_{ab}(0)/\xi_{ab}(0)$ and $\kappa_{ab}(0) = \lambda_{ab}(0)/\xi_c(0) = [\lambda_{ab}(0)\lambda_c(0) / \xi_{ab}(0)\xi_c(0)]^{1/2}$. These equations are used to determine the anisotropic $\lambda_i(0)$ values. Table I gives the GL estimates of $\kappa_i(0)$, $\xi_i(0)$ and $\lambda_i(0)$ for both field orientations (H||*ab* and H//*c*). The anisotropic GL relations require that $\xi_{ab}/\xi_c = \lambda_c/\lambda_{ab}$. The coherence length values $\xi_{ab}$ and $\xi_c$ are 21.3 nm and 12.5 nm respectively, resulting in a ratio of $\xi_{ab}/\xi_c \approx 1.7$, which is in excellent agreement to the $H_{c2}$ anisotropy ratio $\gamma_{anis}(0) = H^{ab}_{c2}(0) / H^{c}_{c2}(0) \approx 1.7$. When comparing these ratios to that of the penetration depth values $\lambda_c/\lambda_{ab} = 25.2nm/66nm \approx 0.4$, it appears that the GL description of this system is invalid. However, the crystals of $Cu_{0.07}TiSe_2$ are thin plates (thickness $d \approx 0.1mm$, area $s \approx 10mm^2$), and demagnetizing fields have a significant effect for field perpendicular to the plates. In what follows, we will take into account the demagnetizing effects in estimating the critical field values, and determine the corresponding values for the affected GL characteristic parameters.

In general the effective field $H^{eff}$ is reduced from the applied magnetic field $H^{app}$ by the demagnetizing field $H_d = N_d M$, where $N_d$ represents the demagnetizing factor, and M is the magnetization of the sample. For a thin plate, $N_d$ is close to 0 when the applied field is parallel to the plate and almost 1 when the field is perpendicular to the plate. In the case of $Cu_{0.07}TiSe_2$, the demagnetizing correction to the applied magnetic field is negligible when H||$ab$. Moreover, as $H_{c2}$ is determined as the magnetic field where the sample enters the normal state, for which M = 0, the measured $H_{c2}$ values are very close to the effective upper critical field values. Thus it appears that, of the critical fields for $Cu_{0.07}TiSe_2$, only the H//$c$ $H_{c1}$ value is significantly affected by demagnetizing effects. Fig. 5 shows the T = 1.8 K H//$c$ magnetization data, as a function of applied field (full symbols, top axis) and as a function of effective field $H^{eff} = H^{app} - M$ (open symbols, bottom axis). The effective value of the lower critical field $H_{c1}$ is determined to be ~50 Oe. This yields a GL coefficient $\kappa_c(0) = 13.4$, with corrected penetration depth values $\lambda_c = 584$ nm and $\lambda_{ab} = 285$ nm, and their ratio $\lambda_c/\lambda_{ab} = 584/285 \approx 2$. This is much closer to the ratio of the coherence lengths $\xi_{ab}/\xi_c$ and the anisotropy ratio $\gamma_{anis}(0) = 1.7$, as required by the GL theory.

In conclusion, we have shown that $Cu_{0.07}TiSe_2$ is a normal type II superconductor, with a transition temperature $T_c = 3.9$ K. The superconducting state properties are anisotropic, characterized by an anisotropy ratio $\gamma_{anis}(0) = 1.7$. In the normal state, the zero-field resistivity measurement yields a RRR around 7, but the overall resistivity values are high, indicative of high scattering mechanisms, possibly reflecting the intrinsically disordered nature of the intercalated copper atoms. Comparison to the layered $NbSe_2$ dichalcogenide superconductor [17], which also shows moderate anisotropy ($H^{ab}_{c2}(0)/H^{c}_{c2}(0) \approx 3$), is supportive of this idea, as the undoped layered $NbSe_2$ has lower in-plane resistivity values. A large $A/\gamma^2$ ratio is also observed for $Cu_{0.07}TiSe_2$, which may be due to the reduced dimensionality of the Fermi liquid or the proximity to the CDW-SC transition. Detailed studies of the Fermi surface in $Cu_{0.07}TiSe_2$ may provide further insight into the anisotropic properties observed in this intercalated layered superconductor.

Table I. Characteristic parameters of $Cu_{0.07}TiSe_2$: critical temperature for superconductivity $T_c$, residual resistivity $\rho_0$ near $T_c$, residual resistivity ratio RRR, upper critical field $H_{c2}(0)$, GL coherence length $\xi(0)$, lower critical field $H^*_{c1}(0)$ (after the demagnetizing correction), anisotropy ratio of $H_{c2}$ $\gamma_{anis}(0)$, GL parameter $\kappa(0)$ and GL penetration depth $\lambda(0)$. The $H//c$ $H_{c1}$ value (before applying the demagnetizing correction) was 13 Oe (see text).

|  | $T_c$ (K) | $\rho_0$ ($\mu\Omega$ cm) | RRR | $H_{c2}(0)$ (T) | $\xi(0)$ (nm) | ‡$H^*_{c1}(0)$ (Oe) | $\gamma_{anis}(0)$ | $\kappa(0)$ | $\lambda(0)$ (nm) |
|---|---|---|---|---|---|---|---|---|---|
| $H\|\|ab$ |  |  |  | 1.238 | 21.3 | 32 |  | 25 | 285 |
| $H//c$ |  |  |  | 0.729 | 12.5 | 53 |  | 13.4 | 584 |
|  | 3.9 | 80 | 6.5 |  |  |  | 1.7 |  |  |

‡$H^*_{c1} = H_{c1} - N_d M$, where $N_d$ represents the demagnetizing factor (see text).

Figure captions:

Fig.1. i∥ab temperature dependent resistivity of $Cu_{0.07}TiSe_2$, with the low temperature part shown in detail in the upper inset; lower inset: low temperatures $\rho(T^2)$ (symbols) with the linear fit (black line).

Fig.2. Field dependent magnetoresistance for (a) H∥ab and (b) H//c at T = 0.35, 0.78, 1.0, 1.3, 1.66, 2.0, 2.5, 3.0 and 3.5 K.

Fig.3. Field dependent magnetization isotherms for (a) H∥ab and (b) H//c at T = 1.8, 2.0, 2.5, 3.0, 3.5 and 3.7 K.

Fig.4. (a) $H_{c2}$ – T phase diagram of $Cu_{0.07}TiSe_2$, with $H_{c1}(T)$ shown in inset. (b) The temperature dependence of the $H_{c2}$ anisotropy ratio $\gamma_{anis}$.

Fig.5. Low field $M_c$ data plotted as a function of applied field $H_{c1}^{app}$ (full symbols) and effective field $H^{eff}_{c1} = H^{app}_{c1} - M$ (open symbols).

Fig.6. A plot of the $T^2$-coefficient of the electrical resistivity A versus the electronic specific heat coefficient $\gamma$ (reproduced from [10]). Solid, dotted and dashed lines represent $A/\gamma^2 = a_0 = 1.0 \times 10^{-5}$ μΩ cm mol² K²/mJ², $A/\gamma^2 < a_0$ and $A/\gamma^2 > a_0$, respectively. Large triangle (▲) corresponds to $Cu_{0.07}TiSe_2$ (present work) and large square (■) corresponds to $Sr_2RuO_4$ ($A_c$) [12].

Fig.1.

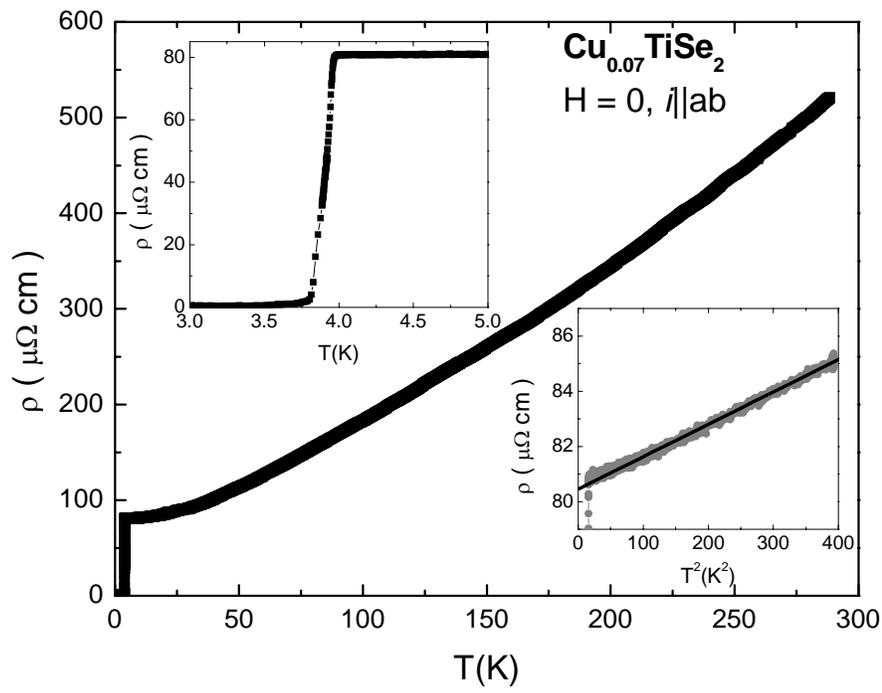

Fig.2.

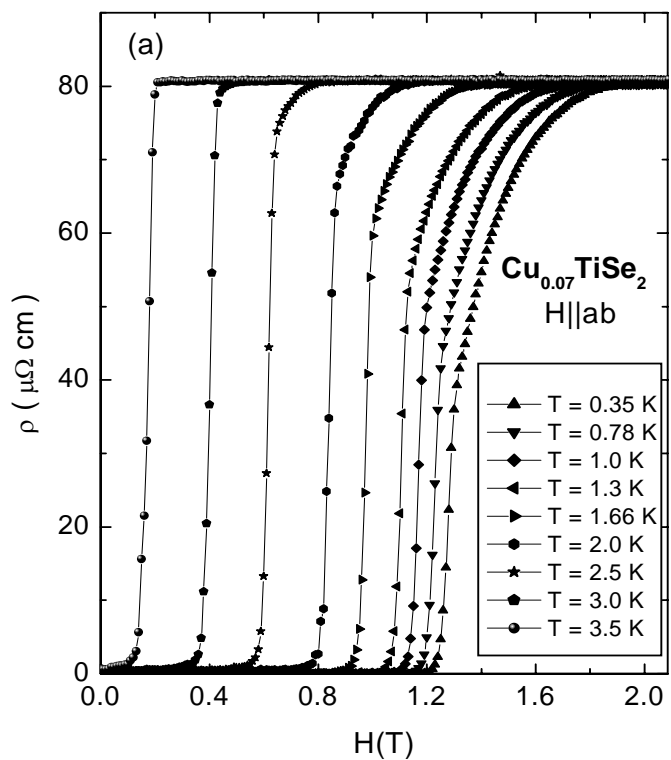 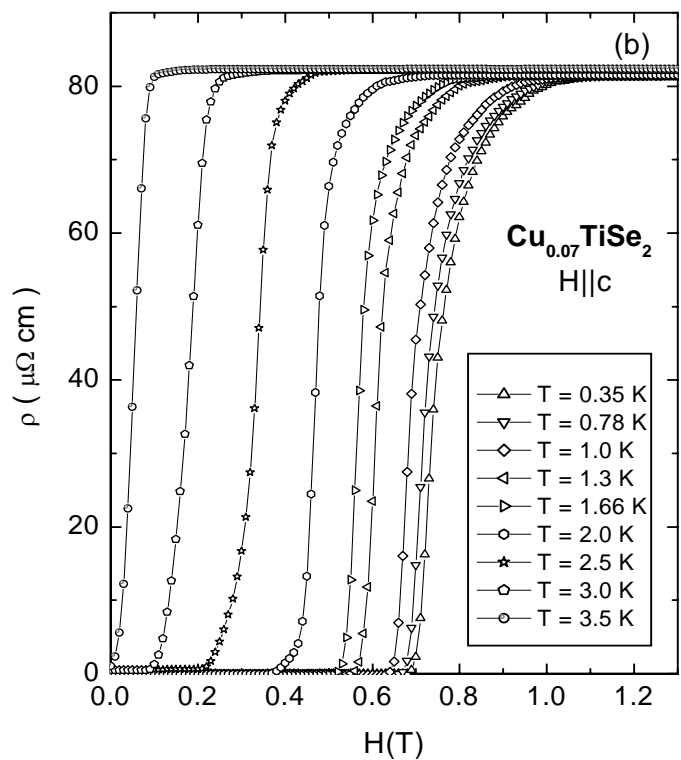

Fig.3.

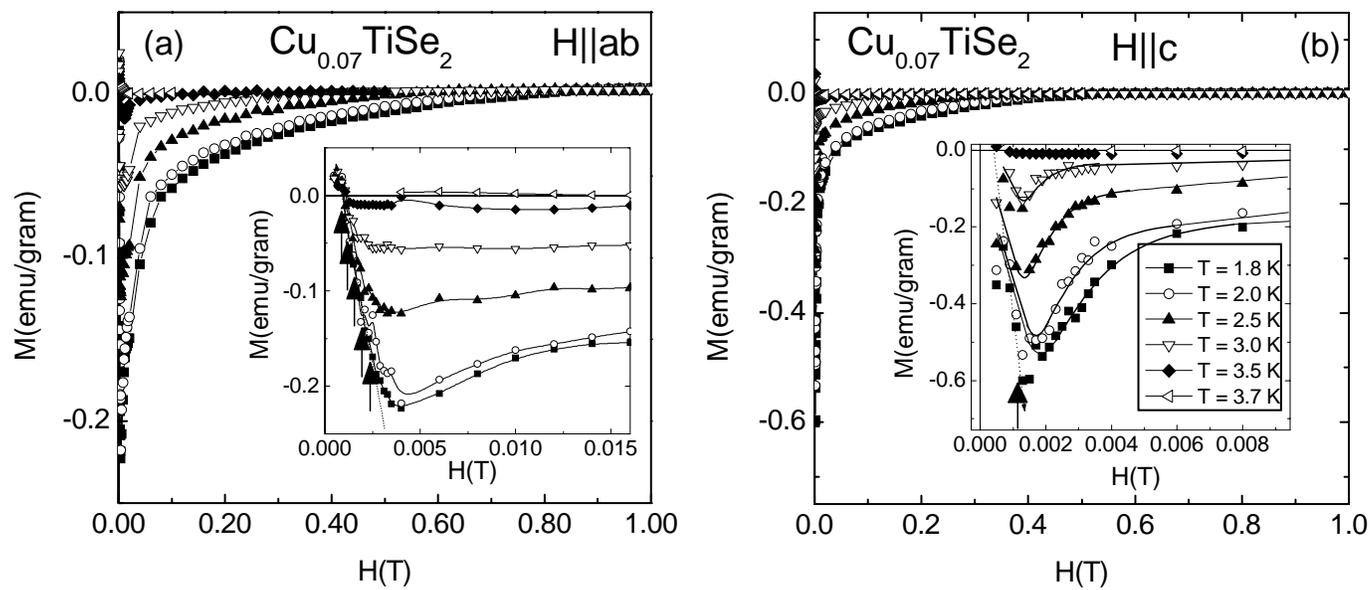

Fig. 4.

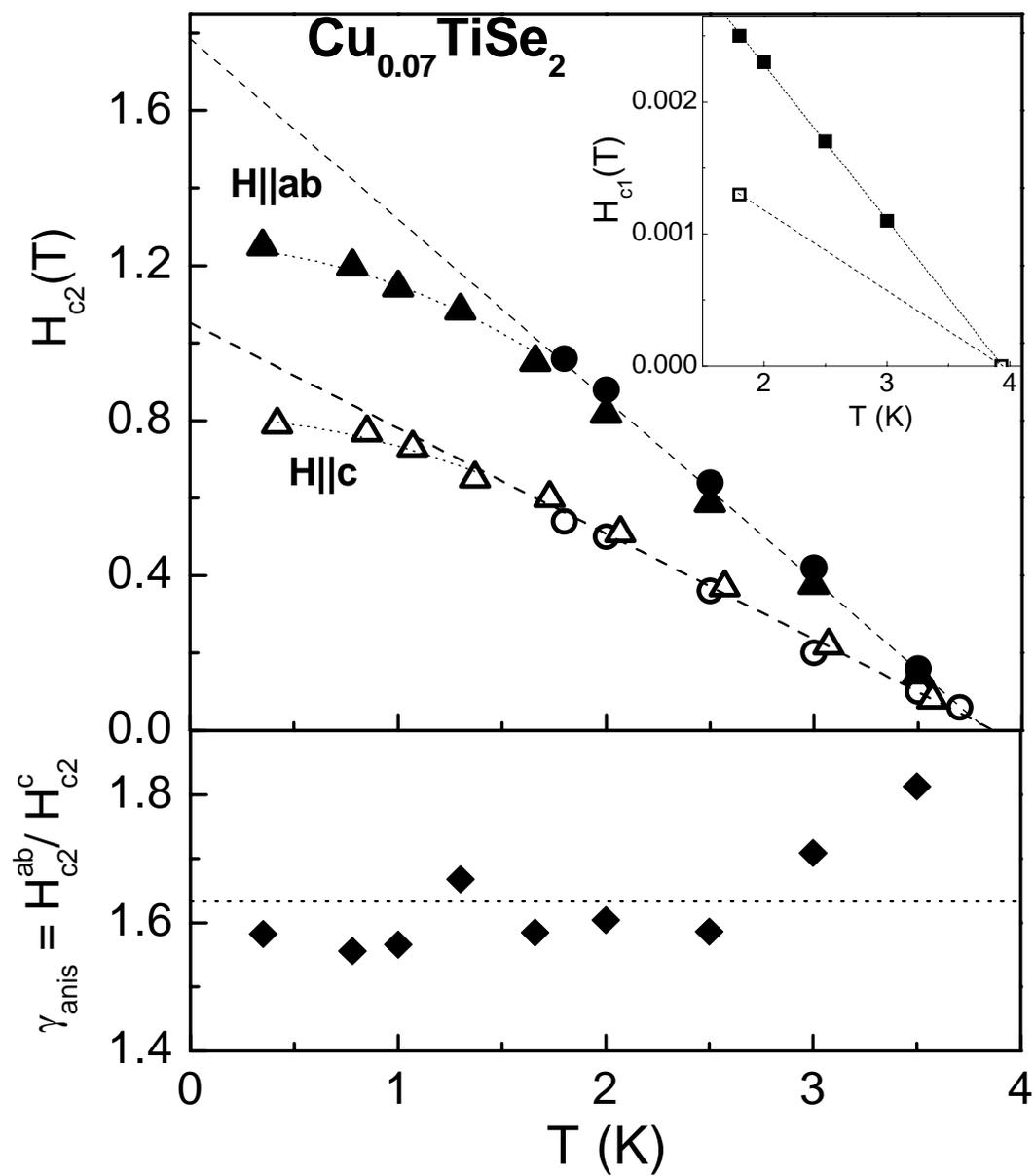

Fig. 5.

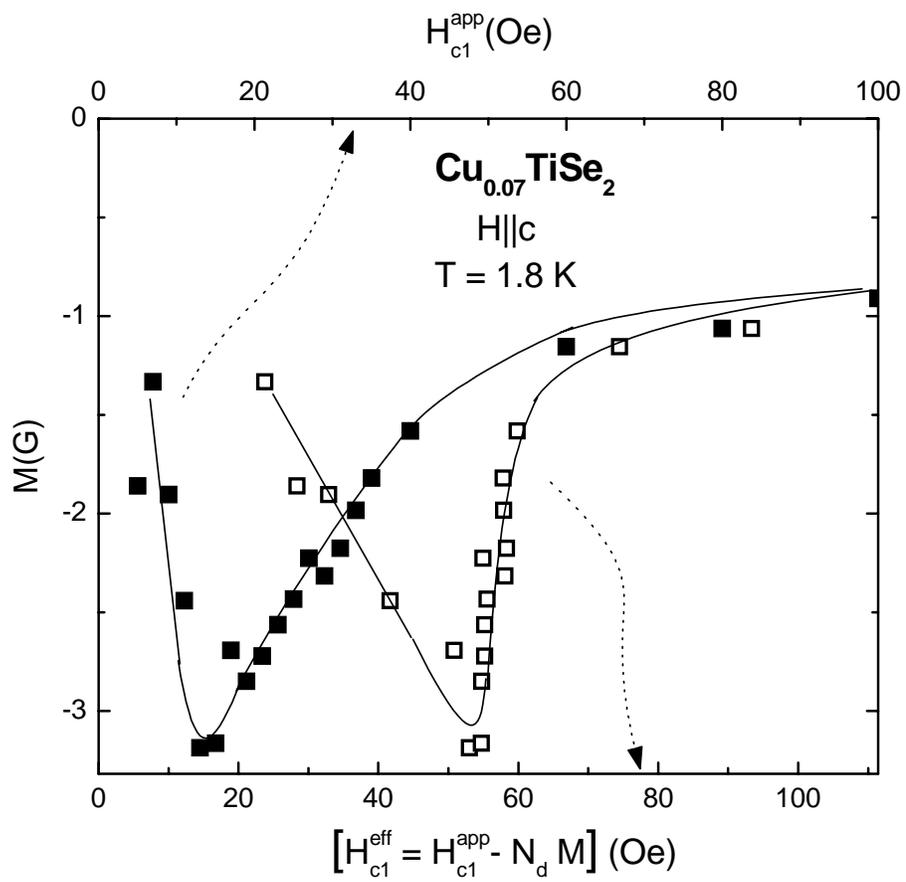

Fig. 6.

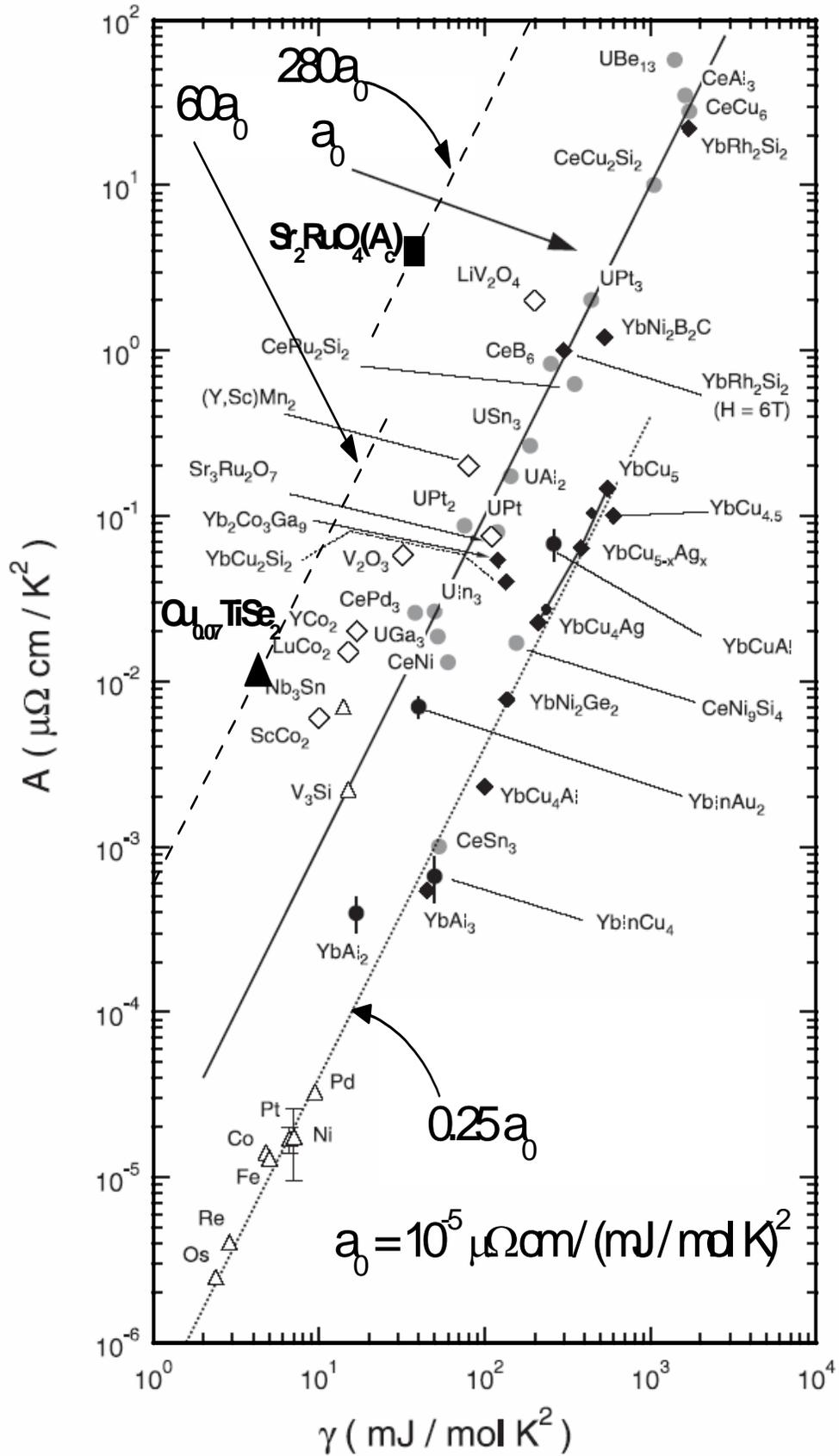